\documentclass[a4paper,11pt]{article}
\usepackage[margin=0.9in]{geometry}
\usepackage{amsmath}
\usepackage{amssymb}
\usepackage{mathtools}

\usepackage{graphicx}
\usepackage{subfigure}
\usepackage{enumitem}  
\usepackage{bbold} 
\DeclareGraphicsExtensions{.png,.pdf}
\usepackage{epsfig}
\usepackage{bm}
\usepackage{color}
\usepackage[colorlinks=true,linkcolor=blue,citecolor=blue,urlcolor=blue]{hyperref}
\usepackage{siunitx}
\usepackage{xcolor}
\usepackage{soul}
\setstcolor{red}
\usepackage[bbgreekl]{mathbbol}
\DeclareSymbolFontAlphabet{\amsmathbb}{AMSb}%
\DeclareSymbolFontAlphabet{\mathbb}{AMSb}
\usepackage{amsfonts}
\usepackage{mathbbol}
\usepackage[utf8]{inputenc}
\usepackage[english]{babel}

\newcommand{\edu}[1]{\textbf{\color{brown} [Eduardo: #1]}}

\usepackage{setspace}

\usepackage[backend=bibtex,url=false,isbn=false,doi=false,articletitle=true,style=chem-acs]{biblatex}

\usepackage{makeidx}

\usepackage{authblk}

\DeclareMathOperator{\Tr}{Tr}

\title{Quantification of electronic and magnetoelastic mechanisms of first-order magnetic phase transitions from first principles: application to caloric effects in La(Fe$_x$Si$_{1-x}$)$_{13}$}
\author[1,*]{Eduardo Mendive Tapia}
\author[2]{Christopher E.\ Patrick}
\author[3,4]{Tilmann Hickel}
\author[3]{Jörg Neugebauer}
\author[5]{Julie B.\ Staunton}

\affil[1]{\textit{Departament de Física de la Matèria Condensada, Facultat de Física, Universitat de Barcelona,
Martí i Franquès 1, E-08028 Barcelona, Catalonia}}
\affil[2]{\textit{Department of Materials, University of Oxford, Parks Road, Oxford OX1 3PH, United Kingdom}}
\affil[3]{\textit{Department of Computational Materials Design, Max-Planck-Institut für Eisenforschung, 40237 Düsseldorf, Germany}}
\affil[4]{\textit{BAM Federal Institute for Materials Research and Testing, 12489 Berlin, Germany}}
\affil[5]{\textit{Department of Physics, University of Warwick, Coventry CV4 7AL, United Kingdom}}
\affil[*]{Author to whom any correspondance should be addressed. \textbf{Email:} e.mendive.tapia@ub.edu}

\date{\today}
\begin{document}

\maketitle
\tableofcontents

\begin{abstract}

La(Fe$_x$Si$_{1-x}$)$_{13}$ and derived quaternary compounds are well-known for their giant, tunable, magneto- and barocaloric responses around a first-order paramagnetic-ferromagnetic transition near room temperature with low hysteresis. Remarkably, such a transition shows a large spontaneous volume change together with itinerant electron metamagnetic features.
While magnetovolume effects are well-established mechanisms driving first-order transitions, purely electronic sources have a long, subtle history and remain poorly understood. Here we apply a disordered local moment picture to quantify electronic and magnetoelastic effects at finite temperature in La(Fe$_x$Si$_{1-x}$)$_{13}$ from first-principles. We obtain results in very good agreement with experiment and demonstrate that the magnetoelastic coupling, rather than purely electronic mechanisms, drives the first-order character and causes at the same time a huge electronic entropy contribution to the caloric response.

\end{abstract}



\section{Introduction}

\label{fundamentals}

Large magnetocaloric and barocaloric effects can come from dramatic changes to the magnetic order of a material when it is subjected to an external magnetic field and/or other mechanical stimuli.
Giant caloric responses for use in solid-state refrigeration can occur around discontinuous (first-order) magnetic phase transitions. A well-understood and extensively observed mechanism driving such sharp transitions is a strong coupling of magnetism with the lattice volume, known as magnetovolume coupling~\cite{PhysRev.126.104,PhysRev.130.1347}. 
There are some materials, however, that exhibit first-order magnetic phase transitions even though their magnetoelastic coupling is weak or even negligible, as reported, for example, in Eu$_2$In~\cite{PhysRevB.101.174437} and Mn$_3$NiN~\cite{PhysRevX.8.041035,PhysRevB.105.064425}.
The fundamental cause here derives from a response of the complex glue of electrons
to the state of magnetic order, which in turn causes a strong feedback on the magnetic interactions and establishes an electronic origin for the magnetic discontinuity.

Purely electronic mechanisms driving first-order transitions have a long and subtle history and there are deeply insightful works such as that on itinerant electron metamagnetism by Wohlfarth and Rhodes~\cite{doi:10.1080/14786436208213848} and, more recently, on electronic effects in La(Fe$_x$Si$_{1-x}$)$_{13}$ compounds by Fujita \textit{et al}~\cite{doi:10.1063/1.370471,PhysRevB.65.014410}. La(Fe$_x$Si$_{1-x}$)$_{13}$ systems, and other derived quaternary compounds, are very promising for applications, being made of relatively abundant, cheap, and non-toxic elements. In particular, the materials have attracted huge attention from the caloric refrigeration research community owing to the giant magnetocaloric and barocaloric effects associated with a first order paramagnetic-ferromagnetic transition near room temperature accompanied by a small hysteresis~\cite{PhysRevB.65.014410,PhysRevB.67.104416,Mañosa2011LaFeSi,MORENORAMIREZ2019406,https://doi.org/10.1002/adma.201000177}. 
This famous magnetic materials class nonetheless is complicated since the changes of the lattice spacings that occur at the transition are rather large. Apparently there is a magnetoelastic coupling which is also playing an important and possibly pivotal role which is intimately connected with the changes to the itinerant electronic structure.
%
%
In this work we describe a predictive first principles approach, based on the disordered local moment 
(DLM) picture~\cite{0305-4608-15-6-018}, that provides both electronic and magnetoelastic contributions to first-order magnetic phase transitions, to examine the subtleties in La(Fe$_x$Si$_{1-x}$)$_{13}$. We obtain a good agreement with experiment regarding both the nature of the transition and also the consequent caloric responses. We quantify both electronic and magnetoelastic effects, their synergy and demonstrate that the first-order character of the transition can be said to be driven by the latter while a large proportion of the caloric effect has an electronic source. 

This work is organised as follows. In section \ref{theory} we describe our theoretical framework for magnetic phase transitions and caloric effects. Section \ref{minimGmag} is devoted to an explanation of how it accounts for the first principles Gibbs free energy of magnetic materials and describes the origin of first-order magnetic phase transitions.
In section \ref{results1} we present results obtained for a specific La(Fe$_x$Si$_{1-x}$)$_{13}$ compound. Our conclusions are given in the final section \ref{summary}.

\section{Caloric effects and {\it ab initio} origin of first-order magnetic phase transitions}
\label{theory}

\subsection{Fluctuating 'spins' and the Disordered Local Moment picture}
\label{DLM}

\begin{figure}[t]
\centering
\includegraphics[clip,scale=0.32]{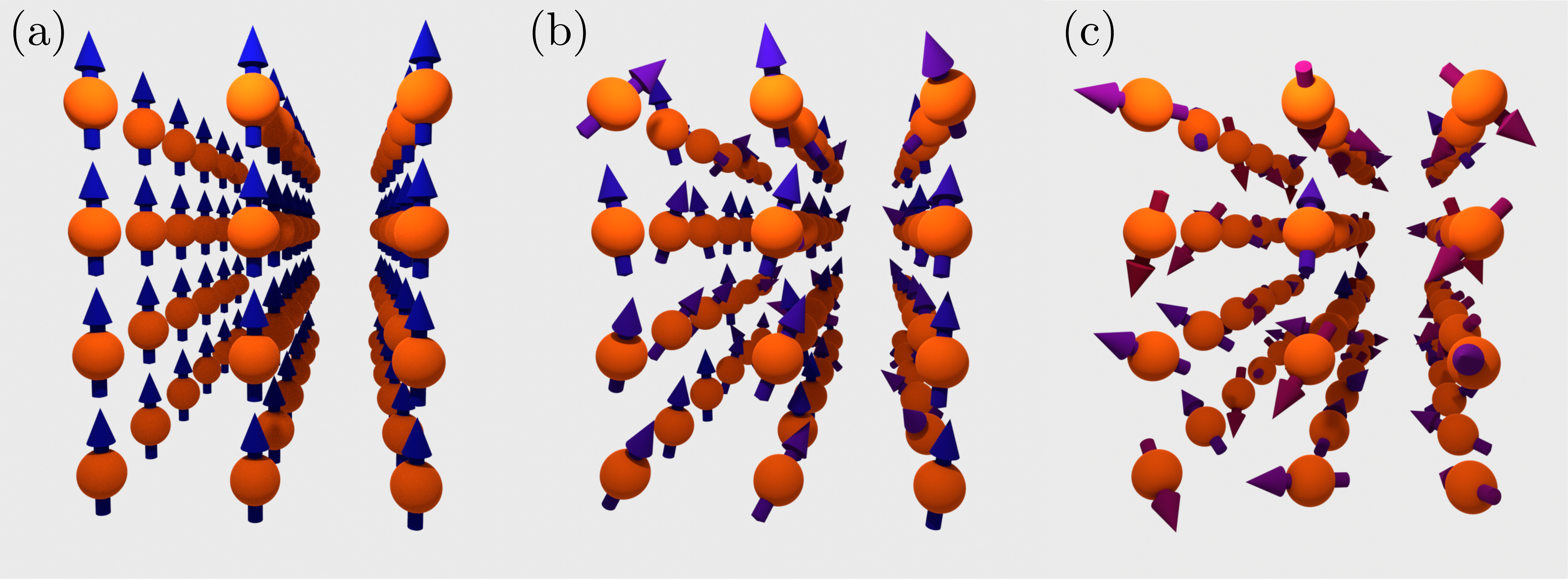}
\caption{
(a) Representation of a ferromagnetic state at zero temperature, described with a value of the magnetic order parameter $m_n=1$ at all magnetic sites, see Eq.\ (\ref{eq_mn}). Panels (b) and (c) show single snapshots of thermally excited local moment configurations for increasing values of the temperature, respectively. Panel (c) shows the fully disordered local moment state where $m_n=0$ at all magnetic sites establishes above the Curie transition temperature $T_c$, while the local moment orientations in (b) form a partially disordered ferromagnetic state below $T_c$ ($0<m<1$).
}%
\label{Fig_DLM}
\end{figure}

The formation of local magnetic moments in a crystalline solid is a process involving spin-correlated electronic interactions which occur over very short (femto-second) timescales, $\tau_\text{elec}$, of the electrons hopping from atomic site to atomic site in the lattice.  The central tenet of the Disordered Local Moment (DLM) theory is to assume that the orientations of these moments are slowly varying degrees of freedom evolving on a much longer time-scale $\hbar/E_{SW} \gg \tau_\text{form} \gg \tau_\text{elec}$, where $E_{SW}$ is a typical spin wave energy and $\tau_\text{form}$ is an intermediate time-scale for the formation of the magnetic moment.
Time averages over $\tau_\text{form}$, therefore, result in a temporarily broken ergodicity labelled by a collection of unit vectors identifying the local moment orientations at each atomic site $n$, $\{\hat{\textbf{e}}_n\}$, which emerge from the many electron interacting system~\cite{0305-4608-15-6-018}. If this assumption holds, the spin-polarisation of the electronic structure in Density Functional Theory calculations can be specified by $\{\hat{\textbf{e}}_n\}$ such that the magnetic moment density, $\bm{\mu}(\textbf{r},\{\hat{e}_n\})$, is constrained to satisfy
\begin{equation}
    \int_{V_n} d^3 \textbf{r} \boldsymbol{\mu}(\textbf{r};\{\hat{\textbf{e}}_n\})
    =\mu_n(\{\hat{\textbf{e}}_n\})\hat{\textbf{e}}_n,
    \label{eq_Muconstr}
\end{equation}
i.e.\ the orientation of the spin polarization within the volume $V_n$ centred at site $n$ is constrained to be along $\hat{e}_n$, $\mu_n$ being its resulting magnitude which can have a dependence on the orientations of the local moments surrounding it.
Figure \ref{Fig_DLM} illustrates this central concept of DLM theory for a ferromagnet by showing different snapshots of constrained local moment configurations at zero and higher temperatures.
Hence, a magnetic Hamiltonian can be established through the dependence of the material's total energy on the local moment orientations,
\begin{equation}
\mathcal{H}_\text{mag}
    = \mathcal{H}_\text{mag,int}(\{\hat{\textbf{e}}_n\},\omega)-\textbf{B}\cdot\sum_n \mu_n\hat{\textbf{e}}_n,
\label{eq_Hmag}
\end{equation}
which defines $\mathcal{H}_\text{mag,int}$ as the Hamiltonian in the absence of an external magnetic field. The `spins' that emerge from itinerant electron metallic systems such as CoMnSi, Fe$_2$P etc are thus the sets of unit vectors specifying the orientations of the local moments, the $\hat{\textbf{e}}_n$'s.
Importantly, the interactions among them can strongly depend on the crystal structure. In this work we focus on the effect of lattice volume changes produced spontaneously and also by the application of a hydrostatic pressure. For convenience we describe them relative to the paramagnetic limit,
\begin{equation}
\omega = \frac{V-V_\text{PM}}{V_\text{PM}},
\label{eq_w0}
\end{equation}
where $V_\text{PM}$ is the unit cell volume obtained from a relaxation in the paramagnetic state.
The nature of the Hamiltonian $\mathcal{H}_\text{mag,int}$ can, therefore, be quite complicated and the sampling of the local moment configurations and the statistical mechanics needs to be carefully carried out, which we address in the following.

\subsubsection{Statistical mechanics of slowly varying local moment orientations}
\label{statistics0}

The description of temperature-dependent properties of materials in terms of the atom-scale magnetic degrees of freedom is provided by $\mathcal{H}_\text{mag}$ through the corresponding calculation of the partition function for the local moment orientations in the canonical ensemble,
\begin{equation}
    \mathcal{Z}_\text{mag}=
    \Tr_\text{mag} \exp\left[-\beta \mathcal{H}_\text{mag}(\{\hat{\textbf{e}}_n\},\omega)\right]
    =
    \int \prod_n\left[d\hat{\textbf{e}}_n\right]
    \exp\left[-\beta \mathcal{H}_\text{mag}(\{\hat{\textbf{e}}_n\},\omega)\right].
    \label{eq_Zmag}
\end{equation}
Here $\beta=\frac{1}{k_\text{B} T}$ is the Boltzmann factor, and $\Tr_\text{mag}$ is the trace over the magnetic degrees of freedom, which takes the form of a product of integrals owing to the continuous nature of each of the  $\{\hat{\textbf{e}}_n\}$.
 The Boltzmann probability distribution associated with the thermal fluctuations of these local moment orientations is given by,
\begin{equation}
    P_\text{mag}(\{\hat{\textbf{e}}_n\},\omega)=\frac{\exp\left[-\beta \mathcal{H}_\text{mag}(\{\hat{\textbf{e}}_n\},\omega)\right]}{\mathcal{Z}_\text{mag}},
    \label{eq_Pmag}
\end{equation}
which leads to the exact magnetic Gibbs free energy of the material by carrying out the appropriate ensemble averages,
\begin{equation}
\begin{split}
   \mathcal{G}_\text{mag}
   & =-k_\text{B} T\ln\mathcal{Z_\text{mag}}
   =\int \prod_n\left[d\hat{\textbf{e}}_n\right] P_\text{mag}(\{\hat{\textbf{e}}_n\},\omega)
    \Big[
    \mathcal{H}_\text{mag}(\{\hat{\textbf{e}}_n\},\omega)
    +k_\text{B} T\ln P_\text{mag}(\{\hat{\textbf{e}}_n\},\omega)
    \Big] \\
   & = U_\text{mag} - TS_\text{mag}.
\end{split}
    \label{eq_G}
\end{equation}
A central task of the DLM approach is to perform the averages over the local moment orientations of different quantities of interest. For example, $U_\text{mag}=\langle \mathcal{H}_\text{mag} \rangle_{P_\text{mag}}$ is the internal magnetic energy, the single-site average of $\hat{\textbf{e}}_n$ is
\begin{equation}
    \textbf{m}_n=\langle\hat{\textbf{e}}_n\rangle_{P_\text{mag}}
    =\int \prod_n\left[d\hat{\textbf{e}}_n\right] P_\text{mag}(\{\hat{\textbf{e}}_n\},\omega)\hat{\textbf{e}}_n,
    \label{eq_mn}
\end{equation}
i.e.\ a magnetic order parameter at site $n$, and the associated magnetic entropy in Eq.\ (\ref{eq_G}) is given by
\begin{equation}
    S_\text{mag} = 
    -k_\text{B}
    \left\langle
    \ln P_\text{mag}
    \right\rangle_{P_\text{mag}}
    =-k_\text{B}\int \prod_n\left[d\hat{\textbf{e}}_n\right] P_\text{mag}(\{\hat{\textbf{e}}_n\},\omega)\ln[P_\text{mag}(\{\hat{\textbf{e}}_n\},\omega)].
    \label{eq_Smag}
\end{equation}

The local moment orientations $\{\hat{\textbf{e}}_n\}$ emerge from the many-electron interactions and as such they have an electronic origin. The DLM approximation splits the trace over all the electronic degrees of freedom $\text{Tr}_\text{elec}$ into the following two pieces~\cite{0305-4608-15-6-018}, the first capturing relatively slowly varying electronic degrees of freedom i.e. the local moment orientations,
\begin{equation}
\text{Tr}_\text{elec}\approx
\text{Tr}_{\{\hat{e}_n\}}\text{Tr}_\text{rest},
\label{eq_Trelec}
\end{equation}
and $\text{Tr}_\text{rest}$ traces over the remaining faster ones.
Eq.\ (\ref{eq_Trelec}) thus reflects a time-scale separation. Such a partition of the trace into sequential parts, $\text{Tr}_\text{rest}$ being performed first for constrained values of $\{\hat{\textbf{e}}_n\}$, implies that $\mathcal{H}_\text{mag}$ is in fact the Gibbs free energy of the faster electronic degrees of freedom,
\begin{equation}
U_\text{mag}=
\langle \mathcal{H}_\text{mag} \rangle_{P_\text{mag}}
\equiv
\langle \mathcal{G}_\text{elec} \rangle_{P_\text{mag}}
=
\langle \bar{E} \rangle_{P_\text{mag}} - TS_\text{elec},
\label{eq_Gelec}
\end{equation}
where $\bar{E}$ is the DFT-based total energy averaged over the faster electronic degrees of freedom, and $S_\text{elec}$ is the associated electronic entropy that has been also averaged over with respect to the local moment configurations$\{\hat{\textbf{e}}_n\}$.

\subsubsection{Computation of caloric effects}
\label{caloric_effects}

It is a common practice to write the total entropy of the material in terms of the following three different parts,
\begin{equation}
    S_\text{tot} = S_\text{vib} + S_\text{mag} + S_\text{elec},
    \label{eq_Stot}
\end{equation}
associated with the fluctuations of vibrational, magnetic, and the remaining electronic degrees of freedom.
The description of caloric effects essentially involves the calculation of entropy changes, both individually and exchanged between these three different components, and of their response to the application of external stimuli. The trace-separation made in Eq.\ (\ref{eq_Stot}) carries fundamental implications.
A careful analysis in relation to Eqs.\ (\ref{eq_Trelec}) and (\ref{eq_Gelec}) shows that Eq.\ (\ref{eq_Stot}) applies to materials where time-scale separations between the corresponding sets of degrees of freedom takes place. 
Particle-hole excitations caused by thermal fluctuations are described by weighting the Kohn-Sham single electron energies of SDFT by the Fermi-Dirac distribution, $f(E;T)=\left(\exp[(E-\nu)/k_\text{B}T]+1\right)^{-1}$, $\nu$ being the chemical potential (Fermi energy $E_\text{F}$ at $T=0$K). The corresponding electronic entropy in Eq.\ (\ref{eq_Gelec}) can then be calculated as
\begin{equation}
    S_\text{elec} (\{\hat{\textbf{m}}_n\},\omega; T) =
    \left\langle
    \int\frac{c}{T}dT
    \right\rangle_{P_\text{mag}}
    \approx
    \frac{\pi^2}{3}k_\text{B}^2T
    \left\langle D(E_\text{F}) 
    \right\rangle_{P_\text{mag}} ,
    \label{eq_SelecSE}
\end{equation}
where the heat capacity, $c$, is the one associated with the internal energy of the single-electron states and $D(E_\text{F})$ is the density of states at the Fermi energy. Although the exact integral form can be formally computed, the approximate expression on the right hand side, derived from a Sommerfeld expansion, is a convenient numerical expression which is easily accessible from DFT calculations. Most importantly, owing to the ensemble averages over the local moment orientations of the density of states at different lattice structures, weighted by $P_\text{mag}$, $S_\text{elec}$ depends on the magnetic order parameters $\{\textbf{m}_n\}$ introduced in Eq.\ (\ref{eq_mn}) and on the relative volume change $\omega$ [see Eq.\ (\ref{eq_w0})].

While the underlying assumption of DLM theory is that the local moment orientations vary much more slowly than the other electronic motions, one can use the Born-Oppenheimer approximation to justify that the vibrations of the heavy nuclei at positions $\{\textbf{R}_n\}$ evolve on a still slower time-scale.
Indeed, Eq.\ (\ref{eq_Stot}) assumes an additional time-scale separation between $\{\hat{\textbf{e}}_n\}$ and $\{\textbf{R}_n\}$. This adiabatic approximation can be used to obtain the vibrational entropy $S_\text{vib}$ as a function of the averages of the faster degrees of freedom~\cite{PhysRevB.76.024309,PhysRevB.78.033102}, calculable within harmonic and quasi-harmonic approximations~\cite{PhysRevB.89.064302,TOGO20151}.
A simpler Debye model can also be employed~\cite{DEOLIVEIRA201089},
\begin{equation}
    S_\text{vib}=k_\text{B}
    \left[
    -3\ln\left(1-e^{-\frac{\theta_D}{T}}\right)
    +12\left(\frac{\theta_D}{T}\right)^3
    \int_0^\frac{\theta_D}{T}
    \frac{x^3}{e^x-1}dx
    \right],
    \label{eq_SvibDebye}
\end{equation}
where $\theta_D$ is the Debye temperature obtainable from experiment or other {\it ab initio} sources~\cite{CHEN2001947}.
%
Results shown in this work for the calculation of caloric effects have been obtained by making use of Eqs.\ (\ref{eq_Smag}), (\ref{eq_SelecSE}), and (\ref{eq_SvibDebye}), $S_\text{elec}$ and $S_\text{mag}$ principally depending on $\{\textbf{m}_n\}$ and $\omega$.
To this end, the Gibbs free energy is computed as a function of $\{\textbf{m}_n\}$ and $\omega$ by carrying out the averages in Eq.\ (\ref{eq_G}) (see section \ref{mean_field}) for different lattice constants. It is then minimized for different values of the temperature and external applied stimuli. We consider the effect of an external magnetic field, $\textbf{B}$, and hystrostatic pressure, $p$.
Isothermal entropy changes can then be directly obtained by applying Eq.\ (\ref{eq_Stot}),
\begin{equation}
\begin{split}
&    \Delta S_\text{iso} (\textbf{B}_0 \rightarrow \textbf{B}_f; T, p) = S_\text{tot}(T,\textbf{B}_f,p) - S_\text{tot}(T,\textbf{B}_0,p), \\
&    \Delta S_\text{iso} (p_0 \rightarrow p_f; T, \textbf{B}) = S_\text{tot}(T,\textbf{B},p_f) - S_\text{tot}(T,\textbf{B},p_0),
\end{split}
\label{eq_deltaSiso_comp}
\end{equation}
for magnetocaloric and mechanocaloric effects, respectively.
Here subscripts $0$ and $f$ indicate initial and final values of the applied fields.
On the other hand, adiabatic temperature changes are obtained by solving the equations
\begin{equation}
\begin{split}
&    S_\text{tot}(T,\textbf{B}_0,p)
     =
     S_\text{tot}(T+\Delta T_\text{ad}(\textbf{B}_0\rightarrow\textbf{B}_f;T,p),\textbf{B}_f,p), \\
&    S_\text{tot}(T,\textbf{B},p_0)
     =
     S_\text{tot}(T+\Delta T_\text{ad}(p_0\rightarrow p_f;T,\textbf{B}),\textbf{B},p_f),
\end{split}
\label{eq_deltaTad_comp}
\end{equation}
where
$\Delta T_\text{ad}(\textbf{B}_0\rightarrow\textbf{B}_f;T,p)$
and
$\Delta T_\text{ad}(p_0\rightarrow p_f;T,\textbf{B})$
are the corresponding thermal responses.

The magnetocaloric effect associated with the second-order phase transition between the paramagnetic and ferromagnetic states in pure gadolinium was one of the first cases for which DLM theory was successfully applied in the context of magnetic refrigeration. The transition temperature $T_c$ as well as isothermal entropy and adiabatic temperature changes were computed in very good agreement with experiment~\cite{Staunton_2014,GschneidnerJr_2005}. Gadolinium's heavier rare earth element next neighbour, dysprosium,  presents a much more complex magnetic phase diagram containing continuous and discontinuous transitions to helimagnetic and fan phases~\cite{Mackintosh1,PhysRevB.71.184410}. DLM theory has demonstrated that the general features of this magnetic phase diagram and consequent magnetocaloric effects have their origin in the feedback on the magnetic interactions among the local moments caused by the response of the mediating valence electrons to different thermal states of magnetic order~\cite{PhysRevLett.118.197202}. Such a purely electronic mechanism is implicit in any metallic system. It underlies the presence of higher than pairwise, multisite, magnetic correlations in the Gibbs free energy, which can generate first order magnetic phase transitions in other materials as shown, for example, in the ordered FeRh alloy~\cite{PhysRevB.89.054427}, Mn-based antiperovskites~\cite{PhysRevX.8.041035,PhysRevB.105.064425} and the Eu$_2$In compound~\cite{PhysRevB.101.174437} that exhibits a near hysteresis-free first-order transition.
In the next sub-section we address the details of how DFT calculations can be used to perform the ensemble local moment orientational averages at the heart of the first principles description of these magnetic phase transitions and magneto caloric effects, the elaboration of multisite correlations and expand on recent computational findings.


\subsubsection{Computationally efficient ensemble averaging of local moment configurations: A mean-field approximation}
\label{mean_field}

Owing to its complex electronic origins in metallic materials, $\mathcal{H}_\text{mag}$, introduced in Eq.\ (\ref{eq_Hmag}), can have a very complicated dependence on the local moment orientations.
A tractable solution is accessible by working with a trial, solvable, Hamiltonian~\cite{0305-4608-15-6-018}
\begin{equation}
\mathcal{H}_\text{tr}= 
-\sum_n\textbf{h}_n\cdot\hat{\textbf{e}}_n,
\label{eq_Htr}
\end{equation}
where $\{\textbf{h}_n\}$ are single-site molecular fields establishing a mean-field theory for the magnetic interactions, sometimes referred to as Weiss fields.
This approach continues by invoking the Peierls-Feynman inequality~\cite{PhysRev.97.660,doi:10.1143/JPSJ.50.1854}, also known as the Gibbs-Bogoliubov inequality~\cite{Isihara_1968,PhysRevB.89.064302}, which provides an upper-bound of the magnetic Gibbs free energy~\cite{PhysRevB.89.054427,PhysRevB.99.144424}:
\begin{equation}
    \mathcal{G}_\text{mag,u}
    =\langle \mathcal{H}_\text{mag,int} \rangle_\text{tr}
    -\textbf{B}\cdot\sum_n \mu_n \textbf{m}_n
    -TS_\text{mag}\geq\mathcal{G}_\text{mag}.
    \label{eq_Gu}
\end{equation}
Here the ensemble averages $\langle\dots\rangle_\text{tr}$ are instead carried out with respect to a Boltzmann probability associated with the trial Hamiltonian,
\begin{equation}
    P_\text{tr}(\{\hat{\textbf{e}}_n\})=\prod_n P_n(\hat{\textbf{e}}_n),
    \label{eq_Ptr}
\end{equation}
where 
\begin{equation}
    P_n(\hat{\textbf{e}}_n) =
    \frac{1}{\mathcal{Z}_\text{tr}}\exp\left[\beta\textbf{h}_n\cdot\hat{\textbf{e}}_n\right],
    \label{eq_Ptrn}
\end{equation}
are single-site probabilities and 
\begin{equation}
    \mathcal{Z}_\text{tr}
    =\prod_n\int d\hat{\textbf{e}}_n \exp
    \left[
    \beta\textbf{h}_n\cdot\hat{\textbf{e}}_n
    \right]
    =\prod_n 4\pi \frac{\sinh(\beta h_n)}{\beta h_n}
    \label{eq_Ztr}
\end{equation}
is the corresponding partition function. The local order parameters and magnetic entropy [see Eqs.\ (\ref{eq_mn}) and (\ref{eq_Smag})] therefore take the form of
\begin{equation}
    \textbf{m}_n=\langle\hat{\textbf{e}}_n\rangle_\text{tr}
    =\int d\hat{\textbf{e}}_n P_n(\hat{\textbf{e}}_n)\hat{\textbf{e}}_n
    =\left[
    \frac{-1}{\beta h_n}+\coth(\beta h_n)
    \right]\hat{\textbf{h}}_n,
    \label{eq_mntr}
\end{equation}
which is the Langevin function, and 
\begin{equation}
   S_\text{mag} 
    = \sum_n S_n
    =-k_\text{B}\sum_n\int d\hat{\textbf{e}}_n P_n(\hat{\textbf{e}}_n)\ln P_n(\hat{\textbf{e}}_n) 
    =k_\text{B} \sum_n
    \left[
    1+\ln\left(4\pi\frac{\sinh(\beta h_n)}{\beta h_n}\right)
    -\beta h_n\coth(\beta h_n)
    \right],
    \label{eq_Smagtr}
\end{equation}
respectively. One can write the corresponding equation of state in this mean-field model by taking the derivative of Eq.\ (\ref{eq_Gu}) and solving $\frac{\partial\mathcal{G}_\text{mag,u}}{\partial\textbf{m}_n} = \textbf{0}$. This results in
\begin{equation}
\left\{
    \textbf{h}_n = \textbf{h}_{n\text{,int}} + \mu_n\textbf{B}
    \right\},
\label{eq_EoSmf}
\end{equation}
where
\begin{equation}
\textbf{h}_{n,\text{int}}
    = -\frac{\partial 
    \langle
    \mathcal{H}_\text{mag,int}
    \rangle_\text{tr}
    }{\partial \textbf{m}_n}
\label{eq_hintmf}
\end{equation}
is the magnetic exchange contribution to the total molecular field and $\mu_n$ is the average magnitude of a local moment on a site.
Increasing $\beta h_n$ to very large values means that $\theta_n$, the angle between $\hat{\textbf{e}}_n$ and $\textbf{h}_n$ is statistically close to zero, which corresponds to the values of the order parameter approaching unity and a low state of magnetic entropy at this site. This is the absolute zero Kelvin limit, while in the high temperature, disordered, limit $m_n = 0$ and the magnetic entropy takes its maximum, $S_n = k_\text{B}\ln(4\pi)$, i.e., the paramagnetic state forms.

The mean-field theory presented above sets up a computationally tractable scheme to calculate the magnetic Gibbs free energy in Eq.\ (\ref{eq_Gu}) by reducing the thermally fluctuating local moment configurations to a numerically affordable and meaningful magnetic phase space given by $P_\text{tr}$. The corresponding ensemble averages over the local moment orientations with respect to $P_\text{tr}$ can be carried out following two different approaches.
The first is based on the so-called coherent potential approximation (CPA)~\cite{PhysRev.156.809,PhysRevB.5.2382,PhysRevB.74.144411}, implementable within the formalism of DFT based on multiple-scattering theory known as the Korringa-Kohn-Rostoker (KKR) electronic structure method~\cite{PhysRev.94.1111,Papanikolaou2002}.
The CPA reproduces the material properties on the average and directly provides the Weiss fields as given in Eq.\ (\ref{eq_hintmf})~\cite{0305-4608-15-6-018}.
Typically self-consistent DLM-DFT calculations are carried out for the paramagnetic state, $T > T_c$, or a fully ordered magnetic state appropriate to zero temperature, to form effective potentials and magnetic fields from the local charge and magnetisation densities.
Further information on this computational method is described elsewhere~\cite{Patrick_2022} and with this implementation DLM-DFT has been successfully applied for a number of years to study temperature-dependent magnetic anisotropy, e.g.~\cite{PhysRevB.74.144411,PhysRevLett.120.097202}, magnetic phase transitions and diagrams of different sorts, e.g.~\cite{HughesNat,PhysRevB.89.054427,PhysRevLett.115.207201,PhysRevLett.118.197202,PhysRevB.95.184438}, and (multi-)caloric effects~\cite{doi:10.1063/5.0003243,PhysRevB.101.174437,PhysRevX.8.041035}.
%
The second approach averages the total energy over $N_\text{MC}$ supercell calculations containing local magnetic moments that are magnetically constrained to follow $P_\text{tr}$~\cite{PhysRevB.105.064425}.
In this cited work we have demonstrated that $N_\text{MC}$ is a computationally accessible small number even for affordable supercell sizes containing a few hundreds of atoms.
It can be implemented in a wide range of DFT codes not restricted to KKR-based ones, and can describe the coupling of magnetism with the crystal structure and consequent lattice relaxations as function of magnetic temperature more accurately.
%
%
%

\subsection{The magnetic Gibbs free energy: First-order magnetic phase transitions}
\label{minimGmag}

One can generally write the average of the magnetic Hamiltonian as
\begin{equation}
    U_\text{mag}= \langle \mathcal{H}_\text{mag,int}\rangle_\text{tr}=
    U^{(0)}
  -\frac{1}{2}\sum_{nn'}U^{(2)}_{nn'}\textbf{m}_{n}\cdot\textbf{m}_{n'} 
  -\frac{1}{4}\sum_{nn'n''n'''}U^{(4)}_{nn'n''n'''}(\textbf{m}_{n}\cdot\textbf{m}_{n'})(\textbf{m}_{n''}\cdot\textbf{m}_{n'''})
    -\dots,
\label{eq_Hmagmn}
\end{equation}
where $\{U^{(2)}_{nn'}, U^{(4)}_{nn'n''n'''}, \dots\}$ are second and higher order coefficients of a Ginzburg-Landau-type expansion in terms of the magnetic order parameters $\{\textbf{m}_{n}\}$.
The  higher order coefficients describe the effect of multisite magnetic interactions on the stability and competition of different magnetic phases at finite temperature~\cite{PhysRevLett.118.197202}.
Terms arising from magnetic anisotropy have not been considered for simplicity.
Most importantly, the sizes and signs of these internal energy coefficients are characteristic of each magnetic material under study. An expression for the Weiss field directly follows by taking the first order derivative to Eq.\ (\ref{eq_Hmagmn})
\begin{equation}
    \textbf{h}_\text{n,int}=
  \sum_{n'}U^{(2)}_{nn'}\cdot\textbf{m}_{n'} 
 + \sum_{n'n''n'''}U^{(4)}_{nn'n''n'''}(\textbf{m}_{n''}\cdot\textbf{m}_{n'''}) \textbf{m}_{n'}
    +\dots
\label{eq_hintmn}
\end{equation}
which can be directly computed. 
The other components of the magnetic Gibbs free energy and their dependence on $\{\textbf{m}_n\}$, i.e.\ the magnetic entropy and the coupling to an applied magnetic field, take a known analytical form in this mean-field model [see Eqs.\ (\ref{eq_Gu}) and (\ref{eq_Smagtr})].  $\mathcal{G}_\text{mag,u}(\{\textbf{m}_n\})$ can therefore be computed, either using the KKR-CPA or supercell approaches, by performing the following steps
\begin{enumerate}
    \item A set of values  $\{\textbf{m}_n\}=\{\textbf{m}_1,\dots,\textbf{m}_N\}$ is chosen, where $N$ is the number of magnetic sites in the magnetic primitive unit cell or the number of magnetically constrained sites in the supercell, respectively.
    
    \item $P_\text{tr}(\{\hat{\textbf{e}}_n\})$ is obtained from $\{\textbf{m}_n\}$ using equations (\ref{eq_Ptr}) and (\ref{eq_mntr}) via mapping through $\{\beta\textbf{h}_n\}$.
    
    \item Information on $U_\text{mag}(\{\textbf{m}_n\})$ is obtained either by computing its first derivative (CPA approach) or its value directly (supercell approach).
    
    \item Steps 1.-3. are repeated for different sets of values of $\{\textbf{m}_n\}$ describing the magnetic states of interest.
\end{enumerate}
At this point such {\it ab initio} data for many sets of $\{\textbf{m}_n\}$ are fit via a linear regression to output a minimal set of internal energy coefficients containing the $\{U^{(2)}_{nn'}, U^{(4)}_{nn'n''n'''}, \dots\}$. This procedure provides the values for a reduced number of magnetic coefficients describing particular magnetic states of interest~\cite{PhysRevB.89.054427,PhysRevB.99.144424,PhysRevB.105.064425}.
The additional dependence of $U_\text{mag}$ on crystal deformation can be calculated by performing the four steps above for different lattice parameters and distortions, which results in the internal energy coefficients as functions of the crystal structure.
Since the internal magnetic energy $U_\text{mag}$ is the only material-dependent term of the Gibbs free energy, the latter is directly given by Eq.\ (\ref{eq_Gu}) once all the terms and dependencies of $U_\text{mag}$ have been obtained following the procedure described here.

In this work we investigate the finite-temperature magnetic properties of La(Fe$_{1-x}$Si$_x$)$_{13}$ compound. The crystal structure of this material is of the type NaZn$_{13}$, which corresponds to the space group $Fm\Bar{3}c$. The internal atomic positions  used in the simulations are those ones observed experimentally ~\cite{Villars2016:sm_isp_sd_1521853} scaled by the relevant lattice parameter.
An important trait of the La(Fe$_{1-x}$Si$_x$)$_{13}$ compounds is that the Fe atoms sit on two non-equivalent sites corresponding to Wyckoff symbols $96i$ and $8b$.  This means that the ferromagnetic state is in fact described in terms of two different magnetic order parameters, $m_{96i}$ and $m_{8b}$.
The results shown in section \ref{results1} are such that $m_{8b}$ is self-consistently obtained along with Weiss fields that fully minimise the Gibbs free energy for a chosen set of values of $m\equiv m_{96i}$, and hence a set of temperatures.  To achieve this we have followed the iterative procedure to compute full equilibrium Weiss fields as done in previous works~\cite{PhysRevMaterials.1.024411,PhysRevB.107.L020401,PhysRevLett.120.097202}.

The internal magnetic energy can thus be expressed as $U_\text{mag}(m_{96i}, m_{8b};\omega)$, where $\omega$ is the relative volume change that we introduced in Eq.\ (\ref{eq_w0}).
In principle, a larger number of DLM-DFT calculations should be done to extract such a full dependence, which also describes non-equilibrium states.
However, the primitive unit cell of  La(Fe$_{1-x}$Si$_x$)$_{13}$ contains 24 and 2 Fe atoms at these Wyckoff positions, respectively. A large contribution to the derivatives of $U_\text{mag}$ might, therefore, come from those made with respect to $m_{96i}$. In this work we have found that this applies to La(Fe$_{1-x}$Si$_x$)$_{13}$ ($x=0.12$) and so have considered the approximation that the internal magnetic energy is a function of $m\equiv m_{96i}$ only. 
In order to validate this approximation we performed again all the calculations presented in section \ref{results1} but setting $m_{8b}=0$, i.e.\ constraining the local moment orientations on these sites to be completely disordered. Following this route we obtained qualitatively similar results. The observed quantitative differences were very small, the transition remains first-order and magnitudes of caloric effects and spontaneous volume changes were almost identical. A significant quantitative difference was that the computed linear magnetovolume coupling of $U^{(2)}$ was 8\% smaller. 
Nonetheless, the role of magnetic order developing on the Fe atoms on the  $8b$ sites deserves a more careful finite-temperature analysis which we plan to perform in future investigations using our supercell DLM approach.

\subsubsection{Electronic and/or magnetoelastic mechanisms}
\label{mechanisms_illustrated}

We illustrate how the analysis of DLM-DFT outputs can be used to investigate first-order phase transitions and quantify their origin by studying the simplest case of a ferromagnet. Since the unit cell of a ferromagnetic phase contains a single crystal position, the magnitudes of the magnetic local order parameters are the same for every site in the lattice. In other words, a single value $m = |\textbf{m}_n|$ is sufficient to specify the state of ferromagnetic order.
We remark that such a scenario also applies approximately to La(Fe$_{1-x}$Si$_x$)$_{13}$, as explained in the former section, as well as to fully compensated antiferromagnetic triangular states, as observed in Mn-based antiperovskite materials~\cite{Matsunami1,PhysRevX.8.041035}.
Inclusion of external forces and spontaneous material responses result in alterations to the unit cell volume, expressed in terms relative to the paramagnetic limit, $\omega=\frac{V-V_\text{PM}}{V_\text{PM}}$ [see Eq.\ (\ref{eq_w0})]. In this situation
Eq.\ (\ref{eq_Hmagmn}) in units of energy per atom becomes
\begin{equation}
    \frac{1}{N_\text{at}} U_\text{mag}= \frac{1}{N_\text{at}}\langle \mathcal{H}_\text{mag,int}\rangle_\text{tr}=
    U^{(0)}(\omega)
  -U^{(2)}(\omega) m^2
  -U^{(4)}(\omega) m^4
    -\text{h.o.},
\label{eq_Hmagmn_simp}
\end{equation}
where $N_\text{at}$ is the number of atoms, and
\begin{equation}
\begin{split}
& U^{(2)}(\omega) = 
    -\frac{1}{2N_\text{at}}\sum_{nn'}U^{(2)}_{nn'}(\omega)\cos\Delta\theta_{nn'} \\
& U^{(4)}(\omega) = 
    -\frac{1}{4N_\text{at}}\sum_{nn'n''n'''}U^{(4)}_{nn'n''n'''}(\omega)\cos\Delta\theta_{nn'}\cos\Delta\theta_{n''n'''}
\label{eq_Hmagmn_simp_U}
\end{split}
\end{equation}
are effective internal energy coefficients containing the effect of the local moment correlations via $\cos\Delta\theta_{nn'}$. These are relative angles between the local order parameters at sites $n$ and $n'$ whose study can be used to extract more free energy coefficients~\cite{PhysRevLett.115.207201,PhysRevLett.118.197202}.
Second and higher order of these coefficients describe the dependence of the single-site Weiss fields against $m$ as given by Eq.\ (\ref{eq_hintmn}),
\begin{equation}
\textbf{h}_{n,\text{int}}=
U^{(2)}(\omega)m
+2U^{(4)}(\omega)m^3
+4U^{(6)}(\omega)m^5
+\cdots .
\label{eq_hvsm_FM}
\end{equation}
The typical parabolic behavior of the energy in the paramagnetic limit ($m\rightarrow 0$) is captured by
\begin{equation}
    \lim_{m\rightarrow 0} U_\text{mag} = U^{(0)}(\omega)=U^{(0)}(V_\text{PM})+\frac{1}{2}\gamma V_\text{PM}\omega^2,
    \label{eq_E0vare}
\end{equation}
where $\gamma$ is the inverse of the compressibility, or bulk modulus, in the paramagnetic state.
For simplicity, we also assume a magnetovolume coupling that is linear in $\omega$. For second, fourth, and sixth order coefficients this means that
\begin{equation}
\begin{split}
  &  U^{(2)}(\omega)\approx
    U^{(2)}_0 + \alpha^{(2)}\omega, \\
  &  U^{(4)}(\omega)\approx
    U^{(4)}_0 + \alpha^{(4)}\omega, \\
  &  U^{(6)}(\omega)\approx
    U^{(6)}_0 + \alpha^{(6)}\omega.
\label{eq_mv_linear}
\end{split}
\end{equation}
The magnetic Gibbs free energy in Eq.\ (\ref{eq_Gu}) therefore is
\begin{equation}
\begin{split}
    \frac{1}{N_\text{at}}\mathcal{G}_\text{mag,u}
    = &
    -\textbf{B}\cdot \mu \textbf{m}
    + p V_\text{PM}\omega
    -TS_\text{mag} \\
 &  +U^{(0)}(V_\text{PM})+\frac{1}{2}\gamma V_\text{PM}\omega^2
  -\left(
  U^{(2)}_0 + \alpha^{(2)}\omega
  \right) m^2
  -\left(
  U^{(4)}_0 + \alpha^{(4)}\omega
  \right) m^4
    -\text{h.o.}.
\label{eq_Gu_simp}
\end{split}
\end{equation}
This result is minimized with respect to $\omega$ to yield
\begin{equation}
\omega=\frac{1}{V_\text{PM}\gamma}\left(
\alpha^{(2)} m^2+\alpha^{(4)} m^4
\right)
-\frac{p}{\gamma}.
\label{eq_wmin}
\end{equation}
The terms in the Gibbs free energy in Eq.\ (\ref{eq_Gu_simp}) can be arranged into different orders of $m$ by using a Taylor expansion for the magnetic entropy,
\begin{equation}
    S_\text{mag}=k_\text{B}\sum_n \left(
    \ln 4\pi -\frac{3}{2}m_n^2-\frac{9}{20}m_n^4 -\cdots
    \right),
    \label{eq_Snexp}
\end{equation}
and Eq.\ (\ref{eq_wmin}) to give
\begin{equation}
\begin{split}
   &  \frac{1}{N_\text{at}}\mathcal{G}_\text{mag,u}
    =
    -k_\text{B} T \ln 4\pi
   +U^{(0)}(V_\text{PM})
   -\frac{V_\text{PM}p^2}{\gamma}
   -\textbf{B}\cdot \mu \textbf{m} \\
 & -\left(
  U^{(2)}_0 - \frac{\alpha^{(2)}p}{\gamma}
  -\frac{3}{2}k_\text{B} T
  \right) m^2
  -\left(
  U^{(4)}_0 +\frac{\left[\alpha^{(2)}\right]^2}{2V_\text{PM}\gamma} - \frac{\alpha^{(4)}p}{\gamma}
  -\frac{9}{20}k_\text{B} T
  \right) m^4
    -\text{h.o.}
\label{eq_Gu_simp2}
\end{split}
\end{equation}
Eq.\ (\ref{eq_Gu_simp2}) can be used to analyse the character of a magnetic phase transition from the paramagnetic state.
If, in the absence of an external magnetic field ($\textbf{B}=\textbf{0}$), such a transition is second-order, i.e., the order parameter changes continuously from $m=0$ to $m\neq 0$ by lowering the temperature, the corresponding transition temperature can be obtained by solving
$\frac{\partial^2\mathcal{G}_\text{mag,u}}{\partial m ^2}|_{m=0}=0$.
This gives
\begin{equation}
    T_\text{tr,sec}=\frac{ 2 }{3k_\text{B}}
    \left(
    U^{(2)}_0 - \frac{\alpha^{(2)} p}{\gamma}
    \right).
    \label{eq_TSOT}
\end{equation}
On the other hand, the transition becomes discontinuous when
\begin{equation}
    \lim_{m\rightarrow 0}\frac{\partial^2\mathcal{G}_\text{mag,u}}{\partial m^2}\Bigg|_{T=T_\text{tr,sec}} = 0^{-}.
    \label{eq_cFOT}
\end{equation}
A first-order character arises, therefore, when the overall fourth order coefficient at $T=T_\text{tr,sec}$ is negative, which gives rise to the following condition
\begin{equation}
    U^{(4)}_0+\frac{\left[\alpha^{(2)}\right]^2}{2V_\text{PM}\gamma}
    +\frac{p}{\gamma}\left(
    \frac{3}{10}\alpha^{(2)}-\alpha^{(4)}
    \right)
    >\frac{3}{10}U^{(2)}_0 .
    \label{eq_cFOT2}
\end{equation}


\begin{figure}[t]
\centering
\includegraphics[clip,scale=0.6]{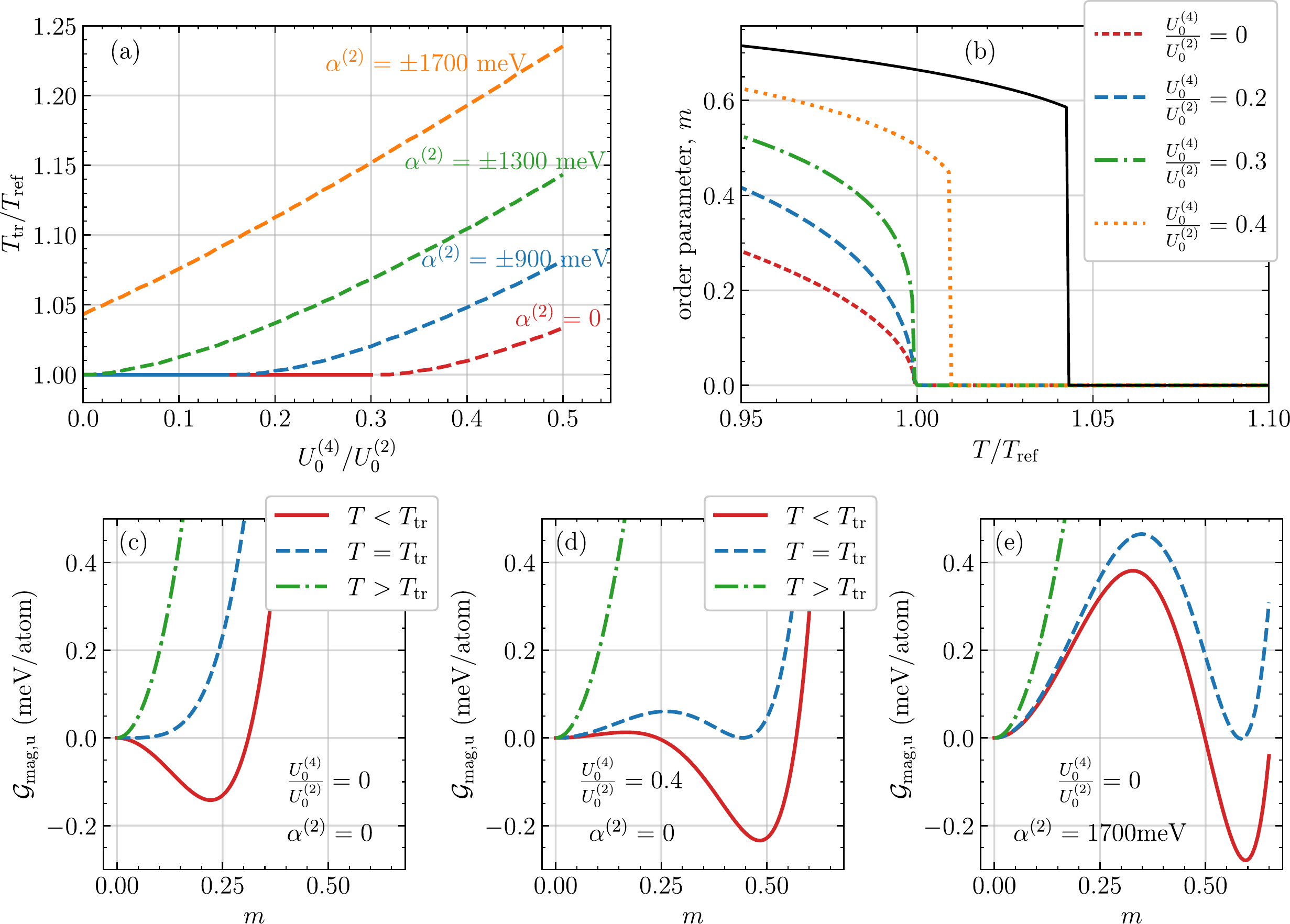}
\caption{
Second- (continuous) and first- (discontinuous) order magnetic phase transitions obtained after minimising Eq.\ (\ref{eq_Gu_simp2}).
(a) Transition temperatures from the paramagnetic ($m=0$) to a (partially-)ordered ($m>0$) magnetic state for different values of $U_0^{(4)}/U_0^{(2)}$ and increasing values of the lowest order of the magnetovolume coupling $\alpha^{(2)}$.
Continuous and discontinuous lines correspond to second- and first-order magnetic phase transitions, respectively.
Representative values of $U_0^{(2)}=190$ meV, $\alpha^{(4)}=0$, $\gamma=200$ GPa, and $V_\text{PM}=11.3$\AA$^{3}$ have been used (approximate for bcc Fe~\cite{PhysRevB.105.064425}), and $T_\text{ref}=\frac{2U_0^{(2)}}{3k_\text{B}}=1470$K is taken as a reference temperature.
(b) Magnetic order parameter against $T$ for different values of the coefficients studied in panel (a). Here $\alpha^{(2)}=0$ for the first four, non-solid, coloured lines, while $U_0^{(4)}/U_0^{(2)}=0$ and $\alpha^{(2)}=1700$ meV apply to the continuous black curve.
Lower panels show the Gibbs free energy below, at, and above $T_\text{tr}$.
Results in panel (d) are also equivalent to those obtained using $U_0^{(4)}/U_0^{(2)}=0$ and $\alpha^{(2)}=1470$ meV.
}%
\label{Fig_panels2}
\end{figure}

Eq.\ (\ref{eq_cFOT2}) demonstrates that a magnetovolume coupling for the pairwise correlations, $\alpha^{(2)}$, always contributes to enhance the first-order character of the magnetic phase transition regardless of its sign. On the other hand, $U^{(4)}_0$ has to be positive to do so.
In the absence of external stimuli, a purely electronic origin for a first-order transition, i.e.\ when $\alpha^{(2)}=\alpha^{(4)}=0$,  occurs if $U^{(4)}_0$ is larger than 30\% of the absolute magnitude of $U^{(2)}_0$. This is shown in Fig.\ \ref{Fig_panels2}(a), where the transition temperature is plotted for different values of $U^{(2)}_0$ and $\alpha^{(2)}$. Here continuous/dashed lines indicate second-/first- order magnetic phase transitions.
These results are obtained after examining the dependence of $m$ on temperature resulting from the minimization of the Gibbs free energy $\mathcal{G}_\text{mag,u}$, which we show in the vicinity of the transition temperature for some illustrative cases in panels (b) and (c-e), respectively.
Indeed, increasing values of $U^{(2)}_0$ and $[\alpha^{(2)}]^{2}$ enhance the first-order character of the transition.
Interestingly, Eq.\ (\ref{eq_cFOT2}) also shows that the application of a hydrostatic pressure can change the character of the transition depending on the signs and relative sizes of the magnetovolume coefficients.

\section{The first-order magnetic transition and calculation of caloric effects in La(Fe$_{1-x}$Si$_x$)$_{13}$ from {\it ab initio} theory}
\label{results1}

La(Fe$_{1-x}$Si$_x$)$_{13}$ is one of the most important and widely studied magnetic materials classes in the field of caloric refrigeration.  There is strong motivation to obtain a fundamental understanding of the first-order magnetic phase transition which is heightened by several intriguing aspects and trends. The Curie transition temperature, varying within the range $T_c=190 - 260$K, decreases together with an enhancement of the first-order character by reducing the content of Si ($x=0.6 - 0.12$), despite the fact that LaFe$_{13}$ does not crystallize in experiment~\cite{PhysRevB.65.014410}. The unit cell volume spontaneously increases by a significant order of magnitude of $1\%$ when the transition is crossed by lowering the temperature from the paramagnetic state, the so-called negative thermal expansion (NTE)~\cite{PhysRevB.65.014410,doi:10.1063/1.1375836}. An itinerant electron metamagnetic origin was directly proposed as the principal mechanism driving the first-order character, ascribed to the observation of how the density of states responds to magnetic stimuli~\cite{PhysRevB.65.014410,PhysRevB.47.11211}. On the other hand, it is reasonable to consider the broadly-invoked magnetoelastic coupling as a source of the discontinuity, which is supported by the substantial spontaneous volume change at the transition. Here we apply DLM-DFT theory to the intermetallic La(Fe$_{0.88}$Si$_{0.12}$)$_{13}$ ($x=0.12$) in order to quantify both electronic and magnetoelastic mechanisms, and we conclude that it is latter mechanism which is driving its first-order character.


All the DLM-DFT calculations presented in this work were performed using the \texttt{MARMOT} code~\cite{Patrick_2022}, an implementation of the DLM picture within the Korringa-Kohn-Rostoker electronic structure method that employs the CPA to simulate different states of magnetic disorder at finite-temperature (see section \ref{mean_field}).
We used an atomic sphere approximation (ASA) together with a scalar-relativistic scheme. The input potentials for \texttt{MARMOT} were generated to simulate the paramagnetic limit. Here Fe sites  are occupied by two magnetic species distinguished by the orientation of their spin moments (‘up’ and ‘down’) with equal concentrations. The local spin density approximation (LSDA) was chosen to treat exchange and correlation effects and the single-site scattering problem was solved using an angular momentum cutoff $l_\text{max}=3$.
The Brillouin zone integration was carried out to very high numerical accuracy, by setting \texttt{MARMOT}'s parameter $\texttt{tolint}$ to $10^{-5}$.
The CPA was also employed to account for the chemical disorder of Si and Fe atoms on the $96i$ sites~\cite{ROSCA201050}, see section \ref{minimGmag} for information on the crystal structure.

\subsection{Construction of the Gibbs free energy: Weiss fields against magnetic order parameters and lattice changes}
\label{Weiss}

\begin{figure}[t]
\centering
\includegraphics[clip,scale=0.65]{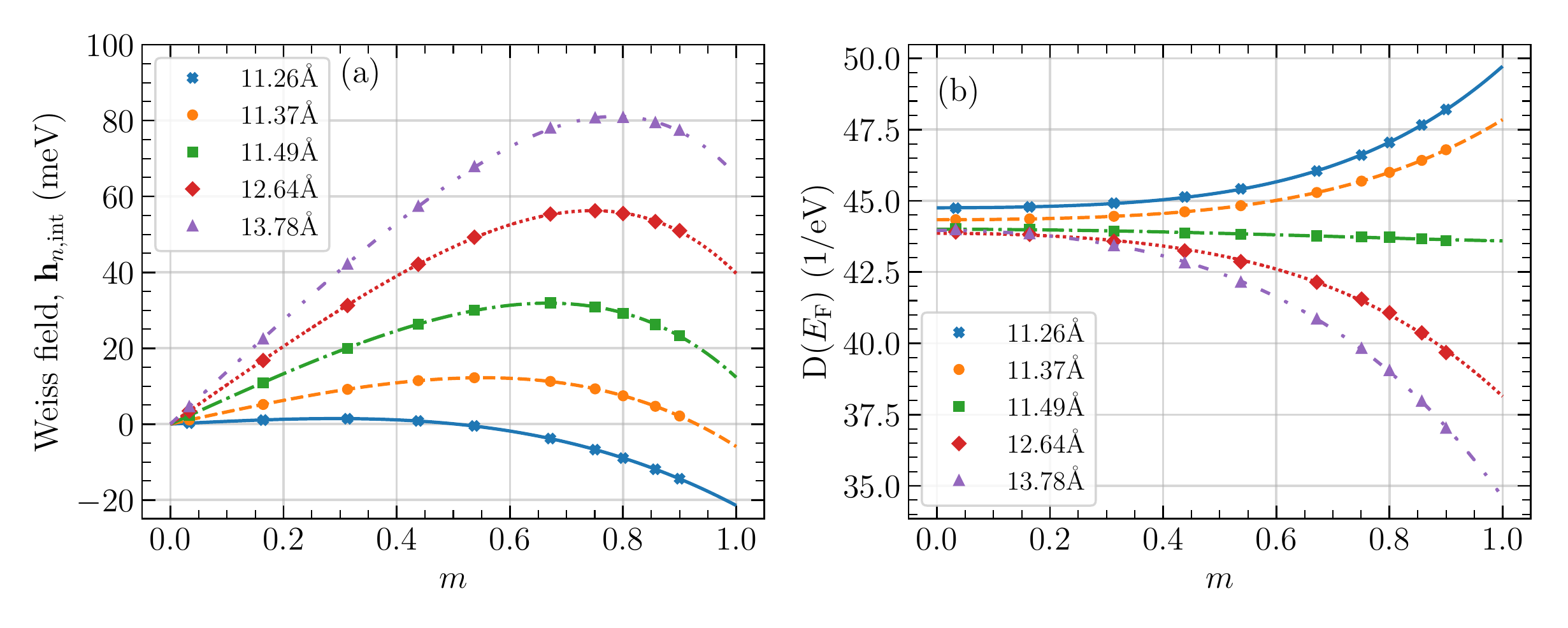}
\caption{
(a) Weiss fields and (b) density of states at the Fermi energy computed for La(Fe$_{0.88}$Si$_{0.12}$)$_{13}$ as functions of the ferromagnetic local order parameter ($m$). Results are shown for different values of the lattice constant. Data points and curves in the graphs correspond to DLM-DFT data and their linear regressions, respectively.
}%
\label{FigLaFeSi_1}
\end{figure}
\begin{figure}[t]
\centering
\includegraphics[clip,scale=0.83]{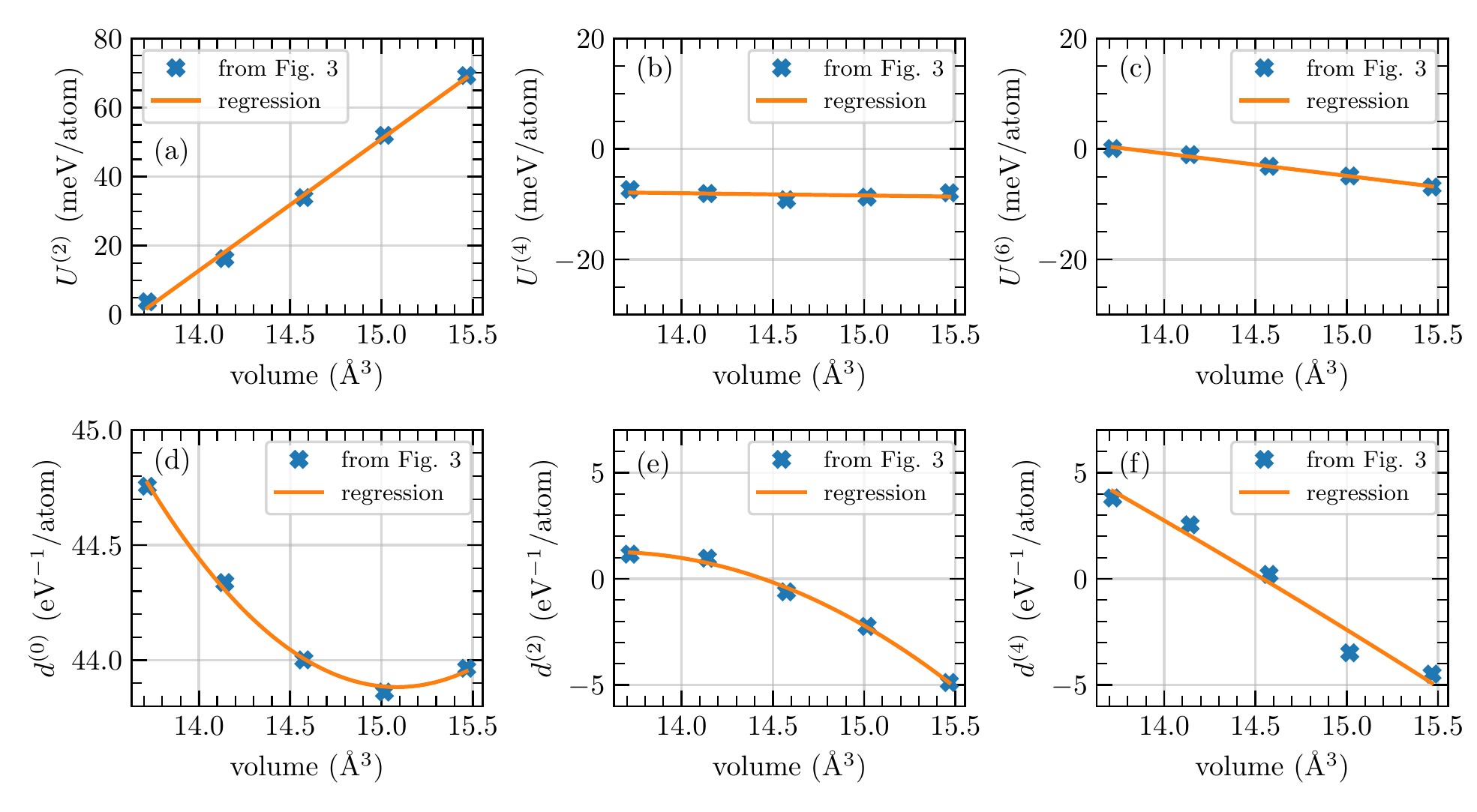}
\caption{
(a,b,c) Internal energy coefficients obtained from a linear regression performed to $\textbf{h}_{n,\text{int}}$, shown in Fig.\ \ref{FigLaFeSi_1}(a), against the volume of the unit cell (per magnetic atom). Continuous curves correspond to the second linear regression performed. Bottom panels (d,e,f) show a similar analysis made for the density of states at the Fermi energy considering our DLM-DFT data in Fig.\ \ref{FigLaFeSi_1}(b) and Eq.\ (\ref{eq_dosEf_fit}).
}%
\label{FigLaFeSi_2}
\end{figure}

Fig.\ \ref{FigLaFeSi_1}(a) shows the internal magnetic field $\textbf{h}_{n,\text{int}}$ computed for La(Fe$_{0.88}$Si$_{0.12}$)$_{13}$, directly given by \texttt{MARMOT}~\cite{Patrick_2022}, as function of ferromagnetic local order parameter ($m$) for different lattice constants, see Eq.\ (\ref{eq_hintmf}). As illustrated in section \ref{mechanisms_illustrated}, the analysis of how $\textbf{h}_{n,\text{int}}$ depends on $m$ and the crystal structure provides the Gibbs free energy of the material, which can be minimized at different temperatures and external fields to obtain the equilibrium states.

In order to achieve this we have performed two consecutive linear regressions. The first is done to $\textbf{h}_{n,\text{int}}$ against $m$ in order to extract the internal energy coefficients $\{U^{(2)}(\omega), U^{(4)}(\omega), \dots\}$ at different values of the lattice constant [see Eq.\ (\ref{eq_hvsm_FM})]. The latter relates to the volume relative to the paramagnetic state, $\omega=\frac{V-V_\text{PM}}{V_\text{PM}}$, where we have considered $V_\text{PM}=14.3$ \AA$^{3}$ per magnetic atom (total unit cell volume of $V_\text{PM}=371.8$ \AA$^{3}$ with $a\approx 11.45$\AA) as a reference value in order to get a closer agreement with experiment. The corresponding regression lines are shown as curves in Fig.\ \ref{FigLaFeSi_1} and the coefficients given in the top panels of Fig.\ \ref{FigLaFeSi_2} as data points. We observe a pronounced linear behaviour for the lowest, leading, second order term $U^{(2)}(\omega)$. This means that there is a substantial magnetovolume coupling despite the fact that $U^{(4)}$ and $U^{(6)}$ barely depend on $V$. 
The second linear regression is applied to extract another set of magnetovolume coefficients, shown in table \ref{table_MVE},  which captures such a linear dependence as given in Eq.\ (\ref{eq_mv_linear}). Indeed, $\alpha^{(2)}$ is the largest and most significant term.

\begin{table}[]
    \centering
    \begin{tabular}{|ccc|}
    \hline
    $\alpha^{(2)}$  & $\alpha^{(4)}$ & $\alpha^{(6)}$ \\
    \hline
    523 meV & -6 meV & -56 meV \\
    \hline
    \end{tabular}
    \caption{\textit{Ab initio} computed magnetovolume coefficients obtained for La(Fe$_{0.88}$Si$_{0.12}$)$_{13}$ from the analysis of the Weiss fields at different values of the magnetic order parameter and lattice constant.}
    \label{table_MVE}
\end{table}


In Fig.\ \ref{FigLaFeSi_1}(b) and bottom panels of Fig.\ \ref{FigLaFeSi_2} we show a similar analysis for the computed density of states at the Fermi energy, $D(E_\text{F})$. Such a calculation is necessary in order to obtain the change of the electronic entropy following Eq.\ (\ref{eq_SelecSE}), which will be used in the next section for the calculation of magnetocaloric and barocaloric effects. We have found that an expansion up to fourth order suffices to model the density of states at $E_\text{F}$:
\begin{equation}
D(E_\text{F})=d^{(0)}+d^{(2)}m^2+d^{(4)}m^4.
\label{eq_dosEf_fit}
\end{equation}
As can be seen in Fig.\ \ref{FigLaFeSi_1}(d,e,f), the second linear regression of these coefficients against the volume, $\{d^{(0)}(\omega), d^{(2)}(\omega), d^{(4)}(\omega)\}$, has required the inclusion of a parabolic magnetovolume coupling.

While the gradient of $\textbf{h}_{n,\text{int}}$ in the paramagnetic limit ($m\rightarrow 0$, described by $U^{(2)}$) gives a major contribution to the value of $T_c$ [see Eq.\ (\ref{eq_TSOT})], the deviation of this linear dependence at higher values of $m$ caused by higher order interactions can generate a first-order character. However, $U^{(4)}$ and $U^{(6)}$ are not only small but negative. This directly implies that such a potential, electronic, source does not contribute positively to the  generation of a discontinuity at the transition, as given by Eq.\ (\ref{eq_cFOT2}). On the other hand, we have found that $\alpha^{(2)}$ is large enough to do so: $\frac{[\alpha^{(2)}]^2}{2\gamma V_\text{PM} U^{(2)}_0}=0.66>\frac{3}{10}$, where $U^{(2)}_0=24\text{ meV}$, and an estimation of the value of the bulk modulus $\gamma=97$ GPa has been directly taken from experiment~\cite{GLUSHKO201940,KAESWURM2017427}. In other words, the first-order phase transition in La(Fe$_{0.88}$Si$_{0.12}$)$_{13}$, and consequent giant caloric effects, which we address in the following, are driven by a strong magnetovolume coupling.

\subsection{Magnetocaloric and barocaloric effects}

\begin{figure}[t]
\centering
\includegraphics[clip,scale=0.65]{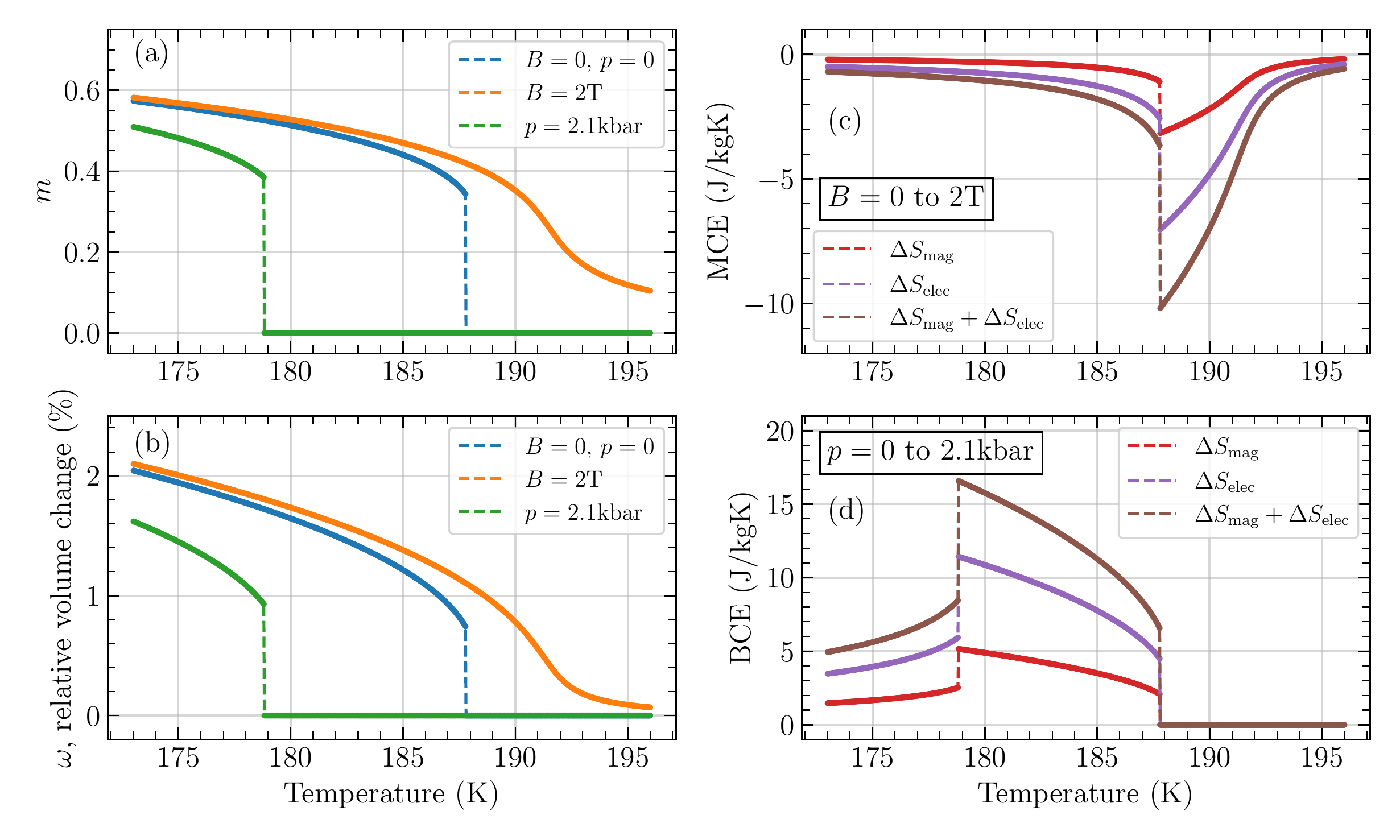}
\caption{
Temperature dependence of (a) magnetic local order parameter and (b) relative volume change, $\omega=\frac{V-V_\text{PM}}{V_\text{PM}}$, computed for La(Fe$_{0.88}$Si$_{0.12}$)$_{13}$ without external stimuli and applying an external magnetic field $B=2$ T and a hydrostatic pressure $p=2.1$ kbar. Corresponding magnetocaloric and barocaloric effects quantified by isothermal entropy changes are shown in the right-hand-side panels (c) and (d), respectively. Their total values are composed of magnetic and electronic contributions.
}%
\label{FigLaFeSi_3}
\end{figure}

We have extracted all the qualitatively and quantitatively relevant internal energy coefficients, $\{U^{(2)}(\omega), U^{(4)}(\omega), \dots\}$, and their dependence on volume, which provides $U_\text{mag}(m, \omega)=-U^{(2)}(\omega)m^2-U^{(4)}(\omega)m^4-U^{(6)}(\omega)m^6$. We can, therefore, now access the Gibbs free energy (per magnetic atom),
\begin{equation}
\mathcal{G}_\text{mag,u}=U_\text{mag}-TS_\text{mag}+\frac{1}{2}\gamma V_\text{PM}\omega^2-B\mu m + pV_\text{PM}\omega .
\label{eq_G_results}
\end{equation}
Note that the DLM-DFT computational analysis that we have done here has yielded more coefficients than those used in Eq.\ (\ref{eq_Gu_simp}), since the latter contained a reduced set for illustrative purposes. Panels (a) and (b) of Fig.\ \ref{FigLaFeSi_3} show the temperature dependence of $m$ and $\omega$ obtained after minimising $\mathcal{G}_\text{mag,u}$ at different values of $T$ with respect to these parameters. The effects of applying an external magnetic field $B=2$T and a hydrostatic pressure $p=2.1$kbar are also plotted. Indeed, we find a first-order phase transition from the paramagnetic to ferromagnetic states. The computed Curie temperature is about $T_c=188$K and the spontaneous volume change at the transition is approximately $\Delta\omega = 0.7$\%, which are in good agreement with experimental measurements.

Table \ref{table_1} summarises a comparison of the results obtained with experiment.
The application of pressure $p$ reduces $T_c$ while enhancing the first-order character, also following experimental trends. The value found is $\Delta T_c / \Delta p \approx -10$ K/kbar, which is very large in accordance with the strong magnetovolume coupling calculated earlier and matches very well the experimental measurement of $[\Delta T_c / \Delta p]^\text{exp} \approx -9$ K/kbar~\cite{PhysRevB.65.014410}. In the right-hand-side panels of the same figure we show the corresponding magnetocaloric and barocaloric effects quantified by the isothermal entropy change. The total value of each of these quantities at the transition is giant as measured experimentally, $|\Delta S_\text{tot}| = 15-20$ J/kgK, being conventional (negative) and inverse (positive), respectively. An important observation is that a large component of such responses comes from the electronic entropy, representing more than 50\% of the total change. This is a consequence of the density of states at the Fermi energy strongly decreasing/increasing in response to the lattice expansion/compression caused when the transition is crossed by applying a magnetic field/pressure, see Fig.\ \ref{FigLaFeSi_1}(b) and Eq.\ (\ref{eq_SelecSE}). A further inspection of Fig.\ \ref{FigLaFeSi_1}(b) reveals that the effect coming from the change of $m$ at the transition, i.e.\ $\Delta D(E_\text{F}) / \Delta m$, changes sign depending on the value of the lattice constant. Importantly, while the large negative thermal expansion is tied to the strong magnetovolume coupling, the sharp change of $m$ is independent of the fact that the origin of the first order character of the transition is not electronic.

\begin{table}[h]
    \centering
    \begin{tabular}{c|ccccccc}
         & $T_c$ (K) & $\frac{\Delta T_c}{\Delta p}$ $\left(\frac{\text{K}}{\text{kbar}}\right)$ & $\Delta\omega$ (\%) & MCE (J/kgK) & BCE (J/kgK) & Refs.\  \\ \hline
        Theory     & 188     & -10 & 0.7 & -10 & +17 & this work \\
        Experiment & 195-208 &  -9 & 1   & -14 &  +9 &  \cite{doi:10.1063/1.1375836,PhysRevB.65.014410,Mañosa2011LaFeSi}
    \end{tabular}
    \caption{Comparison of computed theoretical results for La(Fe$_{0.88}$Si$_{0.12}$)$_{13}$ with experiment. The magnetocaloric (MCE) and barocaloric (BCE) effects are quantified by those isothermal entropy changes observed at the Curie transition temperature that are obtained by applying $B=2$ T and $p=2.1$ kbar, respectively.
    The experimental value for the MCE taken from Ref.\ \cite{doi:10.1063/1.1375836} corresponds to a material with a slightly different composition in comparison with the one studied in this work.
    The experimental value given for the BCE has been taken from Ref.\ \cite{Mañosa2011LaFeSi}, which reports measurements made for a La-Fe-Si-Co compound that exhibits a less pronounced first-order character with a higher transition temperature.}
    \label{table_1}
\end{table}

\subsection{On the origin and nature of the transition and the role of itinerant-electron metamagnetism}
\label{origin}

Our calculations, reported in section \ref{Weiss}, quantify mechanisms for the first-order character of the paramagnetic-ferromagnetic phase transition in La(Fe$_x$Si$_{1-x}$)$_{13}$. We have found that the strong magnetovolume coupling is the principal origin. On the other hand, those complex itinerant electron effects which are entwined with the transverse fluctuations of the local moment orientations, lead to significantly large high order coefficients $U^{(n>2)}$ in our theory. The negative signs of these coefficients, however, act to oppose the occurrence of a first order discontinuity [see Eq.\ (\ref{eq_cFOT2}) and Fig.\ \ref{FigLaFeSi_2}]. While seminal works  carried out for La(Fe$_x$Si$_{1-x}$)$_{13}$ and similar compounds~\cite{PhysRevB.65.014410,FUJITA200662,T_Yokoyama_2001} focus on itinerant electron metamagnetism and claim a crucial role for the coupling of the electronic structure with magnetic order, they also acknowledge the great importance of large magnetoelastic effects and their role in the first-order character of the transition.

In the context of itinerant electron metamagnetism, several authors have investigated the effect of  variations in the sizes $\mu_n$ of the local moments, by means of fixed-spin-moment (FSM) calculations~\cite{PhysRevB.76.092401,FUJITA2012578,Gercsi_2018}. The terms produced by such calculations, involving longitudinal local moment size fluctuations, are fundamentally different from the $U^{(n>2)}$ internal energy coefficients computed in our study based on transverse local moment orientation fluctuations. The application of the FSM approach to La(Fe$_x$Si$_{1-x}$)$_{13}$  has revealed the emergence of several shallow minima in the total energy as a function of the local moment magnitude in the ferromagnetic state ($m=1$) when the lattice is contracted. Such a feature is proposed as a potential factor for the low thermal hysteresis observed in experiment. 

Related itinerant electron physics is captured in the DLM theory in cases where  the size of a local moment on a site is found to depend sensitively on the orientations of the local moments on the sites surrounding it, $\mu_n(\{\hat{\textbf{e}}_n\})$, and hence the extent and nature of magnetic order. Indeed, such insights for La(Fe$_x$Si$_{1-x}$)$_{13}$  could resonate with some features found in early DFT and DLM-DFT studies of face-centered-cubic iron~\cite{PhysRevLett.54.1852, PhysRevLett.56.2096}. Here, as the lattice parameter is decreased, the metal's $T=0$K magnetic state changes from ferromagnetic with a large, $>$2$\mu_B$, magnetisation per atom, to ferromagnetic with a smaller magnetisation, onto antiferromagnetic and eventually non-magnetic. In the paramagnetic DLM state, the size of the local moment on a Fe site, when  averaged over all configurations, collapses to zero for lattices with parameters below a critical value.

\begin{figure}[t]
\centering
\includegraphics[clip,scale=0.67]{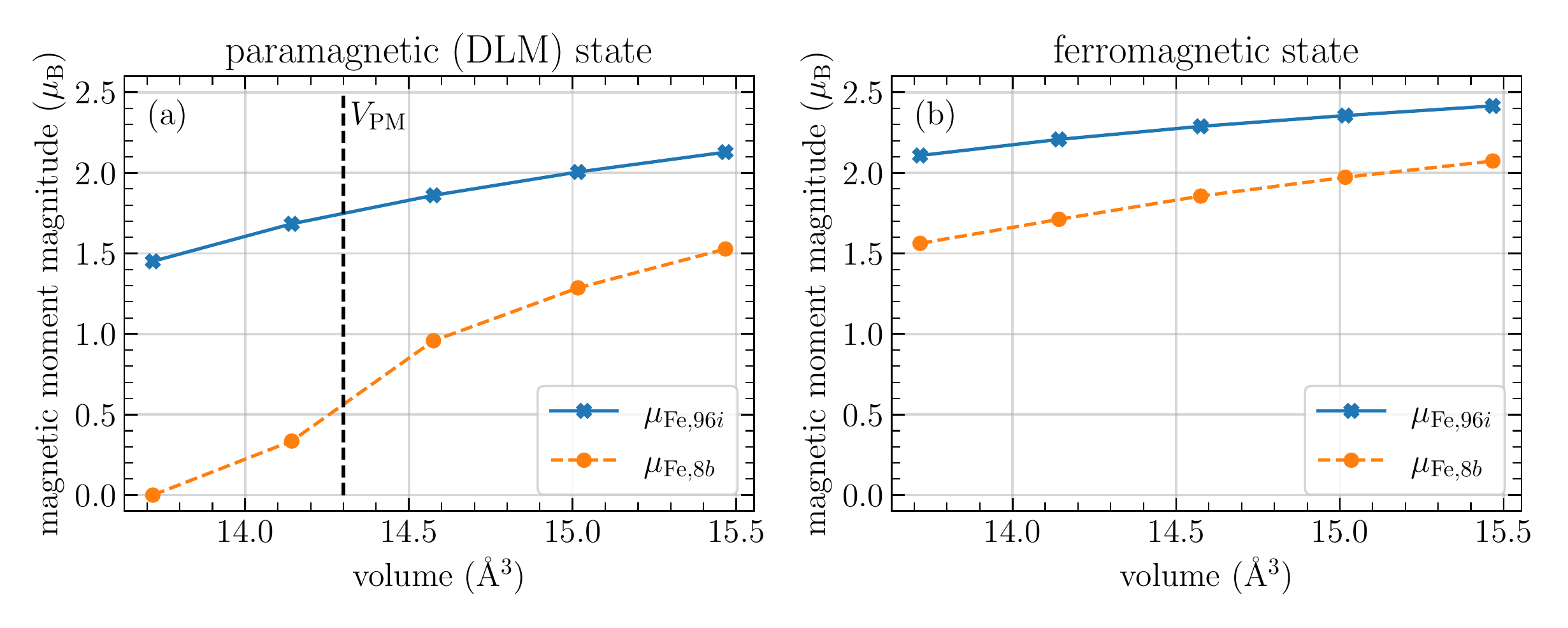}
\caption{
Computed local moment magnitudes at both $96i$ and $8b$ sites in both (a) the disordered local moment (DLM, $m=0$) and (b) the ferromagnetic states as functions of the unit cell volume. The paramagnetic volume in this work, which expands when the ferromagnetic state stabilizes, is indicated by a vertical dashed line.
}%
\label{moments}
\end{figure}

We have investigated similar circumstances for La(Fe$_{0.88}$Si$_{0.12}$)$_{13}$. In figure \ref{moments} we show the computed local moment magnitudes for both the paramagnetic state as well as for the ferromagnetic phase plotted against the unit cell volume. We observe that for the computed paramagnetic volume, $V_\text{PM}$, the local moment of an iron atom at a $96i$ site in the DLM state is reasonably large, $\mu_{\text{Fe},96i}=1.7\mu_\text{B}$, and in agreement with photoemission spectroscopy measurements, $\mu_\text{exp}=1.5 \mu_\text{B}$~\cite{Kamakura2010}. When the first-order transition takes place, the lattice expands and ferromagnetic order develops, both events causing an increase of the local moment sizes. Our calculations fully include the magnetovolume effect but are done throughout using self-consistent DLM potentials produced for the paramagnetic state, $m=0$. In order to investigate the potential influence of fully self-consistent calculations, and hence sensitivity of the moment sizes to magnetic order and unit cell volume, we repeated the calculation of the leading magnetovolume coefficient using potentials from the self-consistent ferromagnetic calculations, $m=1$. The result obtained is $\alpha^{(2)}=415$ meV, which compares reasonably with the DLM outcome given in table \ref{table_MVE} of  $523$ meV and indicates that the effects of the local moments size variations are not overly significant. Further investigations along these lines, however, are needed to fully quantify these aspects.

\section{Conclusions}
\label{summary}

Around a first-order, discontinuous, magnetic phase transition a material can exhibit a large caloric response to the application of a small or moderate external field which results from an abrupt change of the state of thermal order of its local magnetic moments. Analysis shows that such a change originates from a strong dependence of second order coefficients in the material's internal magnetic energy on attributes of the material's crystal structure and electronic glue. There is a consequent shaping of the material's Gibbs free energy with several competing minima and hence a sharp discontinuity in the magnetic order parameters at the transition temperature. Substantial magnetostructural or magnetovolume couplings are well understood sources of first-order transitions in this regard~\cite{PhysRev.126.104}. Purely electronic mechanisms, however, are at an earlier stage of study and have a rich potential for new caloric effects~\cite{doi:10.1080/14786436208213848,doi:10.1063/1.370471,PhysRevB.101.174437}.

In this work we have applied the disordered local moment (DLM) picture in density functional theory (DFT), as a predictive, first-principles tool 
to quantify the extent of and interplay between magneto-structural and electronic sources to the origin of the famous first-order paramagnetic-ferromagnetic phase transition in La(Fe$_{1-x}$Si$_x$)$_{13}$ for $x=0.12$. We have found that the principal origin of such a transition and of the consequent giant magnetocaloric and barocaloric effects is a strong magnetovolume coupling while purely itinerant-electron mechanisms linked to transeverse fluctuations act in opposition to the occurrence of the first-order character. The origin of the transition is linked to large changes of the magnetic order parameter driven by a sharp dependence of magnetic local moment interactions on the value of the lattice constant. The substantial negative thermal expansion computed here and observed in experiment mirrors this fact. The itinerant electron structure, however, which is fundamental to the determining the crystal structure as well as the local moments and magnetic properties, adjusts strongly as the lattice changes and leads to the total isothermal entropy changes with very large magnetic and electronic contributions.
Our work demonstrates that the coupling of the electronic structure in its entirety with the lattice at finite temperature is pivotal for the magnetic and caloric properties of La(Fe$_{1-x}$Si$_x$)$_{13}$. Further investigation of the effects of substituting a small proportion of the Fe atoms with other elements from this perspective show promise for improving and tuning the properties as well as the understanding of this important materials class.


\section*{Acknowledgments}

We are grateful to A.\ Planes for fruitful discussions.
E.\ M.-T.\ acknowledges funding received from the European Union's Horizon 2020 research and innovation programme under the Marie Sklodowska-Curie grant agreement No.\ 101025767, support from MCIN/AEI/10.13039/501100011033 (Spain) under Grant No.\ PID2020-113549RB-I00/AEI,
and computing time provided at the NHR Center NHR4CES at RWTH Aachen University (project number p0020248). The latter is funded by the Federal Ministry of Education and Research, and the state governments participating on the basis of the resolutions of the GWK for national high performance computing at universities.
Presented results have been achieved also using computing time granted by the supercomputer of the Department of Computational Materials Design, operated by the Max Planck Computing and Data Facility in Garching.
This work is supported by the Deutsche Forschungsgemeinschaft (DFG, German Research Foundation) under Projektnummers 405621160 and 405621217. J.\ B.\ S.\ acknowledges support from  EPSRC (UK) Grant No.\ EP/M028941/1 (\href{https://warwick.ac.uk/fac/sci/physics/research/theory/research/electrstr/pretamag/}{PRETAMAG} project).

\printbibliography

\end{document}